%% file: main.tex
\documentclass[11pt, a4paper, logo, copyright]{googledeepmind}

\pdftrailerid{redacted}

\makeatletter
\providecommand\bibentry[1]{\nocite{#1}\fullcite{#1}}
\makeatother

\usepackage{kantlipsum, lipsum}
\usepackage{dsfont}
\usepackage{gdm-colors}
\usepackage[utf8]{inputenc}   
\usepackage{newunicodechar}   
\usepackage{amssymb}          
\usepackage{arydshln}
\usepackage{wasysym,marvosym}
\usepackage{mathtools}
\usepackage{amsmath}

\usepackage{xcolor}
\usepackage{soul}

\soulregister\cite7
\soulregister\ref7
\soulregister\cref7

\usepackage[
  style=authoryear-comp,
  sorting=ynt,
  natbib=true,
  backend=bibtex,
  maxcitenames=2,
  maxbibnames=99,
  uniquelist=false,
  uniquename=false,
  giveninits=true,
  dashed=false
]{biblatex}
\let\cite\textcite

\DeclareFieldFormat{citehyperref}{%
  \DeclareFieldAlias{bibhyperref}{noformat}%
  \bibhyperref{#1}}
\DeclareFieldFormat{textcitehyperref}{%
  \DeclareFieldAlias{bibhyperref}{noformat}%
  \bibhyperref{%
    #1%
    \ifbool{cbx:parens}
      {\bibcloseparen\global\boolfalse{cbx:parens}}
      {}}}
\savebibmacro{cite}
\savebibmacro{textcite}
\renewbibmacro*{cite}{%
  \printtext[citehyperref]{%
    \restorebibmacro{cite}%
    \usebibmacro{cite}}}
\renewbibmacro*{textcite}{%
  \ifboolexpr{
    ( not test {\iffieldundef{prenote}} and
      test {\ifnumequal{\value{citecount}}{1}} )
    or
    ( not test {\iffieldundef{postnote}} and
      test {\ifnumequal{\value{citecount}}{\value{citetotal}}} )
  }
    {\DeclareFieldAlias{textcitehyperref}{noformat}}
    {}%
  \printtext[textcitehyperref]{%
    \restorebibmacro{textcite}%
    \usebibmacro{textcite}}}

\usepackage{bbding}
\usepackage[utf8]{inputenc} 
\usepackage[T1]{fontenc}    
\usepackage{hyperref}       
\usepackage{url}            
\usepackage{booktabs}       
\usepackage{nicefrac}       
\usepackage{microtype}      
\usepackage{amsmath}
\usepackage{graphicx}
\usepackage{tablefootnote}
\usepackage{multicol}
\usepackage[nameinlink]{cleveref}
\usepackage{bbm}
\usepackage{multirow}
\usepackage{soul}
\usepackage{float}
\usepackage{wrapfig}
\usepackage{blindtext}
\usepackage{tablefootnote}
\usepackage{amsfonts}
\usepackage[flushleft]{threeparttable}
\usepackage{colortbl}
\usepackage{mathtools,amssymb}
\usepackage{bm}
\usepackage{makecell}
\usepackage{caption}
\usepackage{capt-of}
\usepackage{array}
\usepackage{calc}      
\usepackage{caption}   
\usepackage{subcaption}  
\usepackage{xcolor,colortbl}
\usepackage[bottom]{footmisc}

\usepackage{tcolorbox}

\definecolor{mydeepgreen}{RGB}{0, 100, 0}

\input{commands}

\graphicspath{{figures/}}

\title{KUAISHOU Explorer LLM-Rec Challenge 2026: Reasoning Generative Recommendation}

\renewcommand{\today}{}

\author{\large OneRec Team}

\begin{abstract}
Generative recommendation, has been attracted a surge of attentions in industrial and academic research community, towards to build more smart system to build next-generation recommender.
Under the significant developing wave of large language model, our team have been developed Semantic ID based OneRec/OneRec-V2.
These models have been widely deployed in production and demonstrate the scaling potential of the autoregressive next-item prediction paradigm for industrial recommender systems.
Building on the success of OneRec, we further explored a series of models, including OneRec-Think, OpenOneRec, and OneReason, that connect item Semantic IDs with natural language in a unified representation space and seek to unlock the potential of natural-language chain-of-thought (CoT) reasoning for recommendation.
However, our preliminary works found that introducing reasoning CoT does not always improve the recommendation performance.
To address this issue, OneReason strengthens the semantic alignment between items and language, introduces structured template-based supervision for interest reasoning, and applies advanced reinforcement learning techniques to make reasoning more beneficial to recommendation.
As a frontier topic to building recommendation foundation models, we believe this topic has significant research value and hope to encourage more researchers to explore it together. 
To this end, together with the SIGIR 2026 community, we organized the KUAISHOU Explorer LLM-Rec Challenge 2026: Reasoning Generative Recommendation.
The challenge comprises four task categories, aiming to develop a unified model that understands items, users, recommendation, and the world knowledge.
Since the challenge is finished at 2026 September, this report introduces the challenge design, describes the open-source data, evaluation benchmarks in detail, and summarizes the leading solutions and their key insights.
To facilicate the research of LLM and RecSys connection, all the corresponding model checkpoint, reasoning SFT data, non-reasoning interaction sequence, and evaluation benchmarks are opensourced.
\\
\\
Challenge website: \url{https://ks-llmrec.streamlake.com/}. 
\end{abstract}

\begin{document}

\maketitle

\begin{figure}[H]
  \centering
  \includegraphics[width=\textwidth]{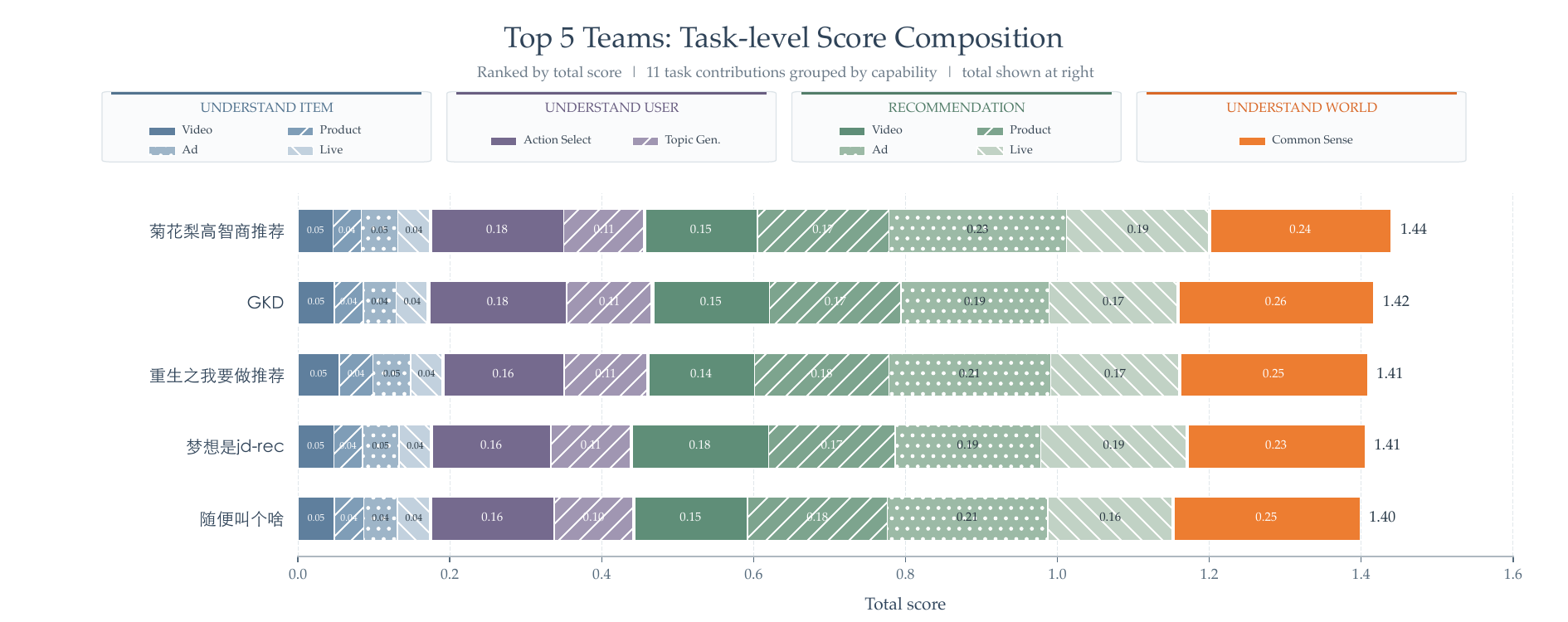}
  \caption{Task-level score composition of the top five teams.}
  \label{fig:top5-team-score-composition}
\end{figure}

\section{Introduction}
\label{sec:introduction}

Generative recommendation~\cite{rajput2023recommender, deng2025onerec, zhang2026onemall, wang2026onelive} formulates the prediction of users' subsequent interactions as a sequence generation problem, opening new direction for scaling recommendation models in the large model era~\cite{2018gpt1, 2019gpt2, 2020gpt}.
In industrial deployment, the OneRec family has advanced this paradigm with Semantic IDs~\cite{deng2025onerec, zhou2025onerec}, and the OneReason~\cite{team2026onereason} has further connect Semantic ID with natural language model.
In this way, the recommendation models have begun to use natural language to understand item content, analyze user interests, and think before making recommendations, and further understanding user interests and how they evolve by combining content semantics with behavioral evidence.

Introducing reasoning~\cite{guo2025deepseek}, however, does not necessarily improve recommendation performance.  
Our preliminary works~\cite{zhou2025openonerec,liu2025onerec} found that models could generate recommendation chains of thought (CoTs)~\cite{lightman2024lets}, yet their thinking mode could still underperform direct generation. 
OneReason addresses this problem through stronger alignment between Semantic ID and language token, structured supervision for interest reasoning, and advanced reinforcement techniques. 
In our findings, the recommendation-oriented CoT designing could significant enhance the pretrainedSemantic-alignment LLM performance, and further could enhance model roll-out beneficial reasoning trajectories for RL/RFT~\cite{ahmad2025opencodereasoning}.
At Kuaishou, OneReason has been deployed in advertising services, which further validates building the recommendation foundation model has significant research value for both academic and industry community.
As a frontier topic to build next-generation recommender, we ackonwledge that we only conduct a small step in the borader research area.
To accelerate the progress in this direction, we with SIGIR 2026 community to organize the KUAISHOU Explorer LLM-Rec Challenge 2026: Reasoning Generative Recommendation, hope to encourage more researchers to explore this topic together.

In recent years, some other famous recommendation competition challenges provide valuable data and evaluation settings for research at industrial scale. 
The TAAC 2025~\cite{pan2026tencent, cao2025onepiece} releases a large-scale advertising interaction sequences with item side multi-modal embedding, focusing on predicting the next advertisement a user will click on or convert, to explore generative recommendation paradigm.
The TAAC $\times$ KDD Cup 2026 challenge~\cite{zhang2026field,lin2026unidot} studies joint modeling of behavioral sequences and non-sequential features to predict conversion probabilities for target item, to explore ranking model feature-crossing mechanism.
Different with them that focusing on iterating existing recommendation model, we want to connect the recommendation system and large language model together, to explore how RecSys can understand content and users and use these capabilities to improve recommendation performance.

To this end, we host the \textbf{KUAISHOU Explorer LLM-Rec Challenge 2026}, which is built on OneReason-0.8B-pretrain/OneReason-8B-pretrain tuned from Qwen-3 series LLM~\cite{yang2025qwen3} with more then 500 Billion Tokens pre-trained.
To help participants explore the reasoning recommendation, the challenge provides 792,797 task-formatted instruct supervised fine-tuning (SFT) data, while 412,438 of them have CoT trajectory.
In addition to these resources, the challenge provides 500,000 anonymous user records spanning short video, e-commerce, advertising, and live streaming, together with SIDs, content descriptions, and category information linked through item identifiers.
Participants can use the supplied examples or organize user histories and construct their own instruction data for SFT and subsequent optimization.
In evaluation, our goal is to bring four capabilities together in one model: understanding items, understanding users, understanding recommendations, and understanding the world. 
This report introduces the competition setting and its OneReason foundation, describes the data and evaluation tasks, and reviews leading participant approaches and the lessons learned from them.

\section{Challenge Setting}
\label{sec:challenge-setting}

\subsection{Problem Definition Overview}
\label{sec:problem-definition}

This challenge goal is to train a single model based on OneReason to perform the following four types of tasks:

\paragraph{Understanding Item.}
In OneReason, each item is formulated as Semantic IDs~\cite{liu2026crem}:
$$
\text{<|}\texttt{domain}\text{\_begin|><s\_a\_x><s\_b\_y><s\_c\_z>},
$$
where \texttt{domain} is one of \texttt{video} (video), \texttt{prod} (product), \texttt{ad} (ad), or \texttt{living} (live) for our four recommendation scenarios, and each semantic ID space within 8192~\cite{luo2025qarm}.
In the preliminary round, the model generates caption from Semantic IDs; in the semi-final round, it generates caption from Semantic IDs. The task covers four domains: short video, e-commerce, advertising, and live streaming.
\begin{equation*}
\begin{aligned}
\texttt{Item Caption}
&\xrightarrow{\text{Preliminary round}}
\texttt{Item Semantic ID} \\
\texttt{Item Caption}
&\xleftarrow{\text{semi-final round}}
\texttt{Item Semantic ID}
\end{aligned}
\end{equation*}

\paragraph{Understanding User.}
Given a user's chronological interaction history and a specified interest topic, the model performs two tasks: selecting some partial interactions in history that are relevant to given topic, and generating an interest-evolution chain from these interactions to describe how the user's needs develop and change around the topic.

\paragraph{Understanding Recommendation.}
The model uses users interaction histories across multiple domains to predict the items that the user is likely to interact with next in a specified domain~\cite{cao2023towards}. Each example specifies one target domain, e.g., short video, e-commerce, advertising, or live streaming, and the model generates candidate items' semantic IDs.
In evaluation stage, we set a two step prediction workflow: (i) using the thinking/non-thinking mode to predict the Semantic ID candidates in Pass@32 setting separately, (ii) merge the thinking and non-thinking prediction results to calculate the final score.
We design the metric aims at the thinking mode could generate more different and diverse user interested items that non-thinking is not able to predict.

\paragraph{Understanding World.}
Given a general knowledge question~\cite{li2024cmmlu,wei2023cmath, huang2023c} and a set of answer options, the model selects the correct option. This task assesses the general knowledge retained after training on recommendation tasks.

\subsection{Competition Stages Overview}
\label{sec:competition-stages}
The challenge consists of a preliminary round, a semi-final round, and a onsite in-person final round.

\paragraph{Preliminary round.}
The preliminary round is started in July 2026. Participants started from OneReason-0.8B-pretrain~\footnote{\url{https://huggingface.co/OpenOneRec/OneReason-0.8B-pretrain-competition}} and trained their models using the provided user interactions (500,000 user sequence), and partial seed SFT data (0.05B Tokens). Evaluation covered item understanding, user understanding, cross-domain recommendation, and general knowledge. The item-understanding task required generating Semantic IDs from a content description. The top 50 teams advanced to the semi-final round.

\paragraph{Semi-Final round.}
The semi-final round took place in August 2026. The base model was upgraded to OneReason-8B-pretrain~\footnote{\url{https://huggingface.co/OpenOneRec/OneReason-8B-pretrain-competition}}, with full seed SFT training data (0.3B Tokens) and computing resources provided for this stage (4 $\times$ 80GB premier GPUs). The item-understanding task changed to Semantic ID to Caption generation with LLM judge, assessing the model's ability to recover content semantics from item codes. Participants continued to improve their models across the four task types and submitted technical reports describing their data processing, training methods, and experimental results.

\paragraph{Onsite Final round.}
According to leaderboards score, the organizers reviewed participants' code, training data, and conducted seriously reproducibility checks to determine the 20 finalist teams. The final round took place in September 2026, when the finalists presented their methods, experimental results, and key findings. The champion/runner-up/third team will earn 400,000/200,000/100,000 CNY, and the top 4-10 teams will award 30,000 CNY, and four innovation special awards (20,000 CNY).
Last but not least, the finalist have the (K-star) full-time or internship job chance to join Kuaishou recommendation group.

\section{OneReason-Pretrain Baseline}
\label{sec:onereason-foundations}
The challenge starts from the OneReason-pretrained model, which only align the item Semantic IDs and natural language within a single auto-regressive model.
Through semantic alignment pretraining, it connects item content, user behavior, and language, providing a solid foundation for subsequent user understanding and recommendation learning. This section describes the base model's item representations, vocabulary expansion, and semantic alignment pretraining at multiple levels of granularity.

\subsection{Item Representation and Vocabulary Expansion}
\label{sec:item-representation}

OneReason uses precomputed semantic IDs (SIDs) to represent items from the short-video, e-commerce, advertising, and live-streaming domains. A separate item tokenizer converts item content into SIDs, which are then used to construct training data for the language model.

The item tokenizer is built through content encoding and discrete quantization. First, a multimodal encoder compresses content such as cover images, video frames, text, and audio into dense vectors. To preserve item semantics, the encoder is jointly trained with a language decoder that reconstructs item descriptions: each item vector is provided to the decoder as a soft prompt, from which the corresponding description is generated. RQ-KMeans then applies three levels of residual quantization to the vectors, with 8,192 codewords at each level. Each item SID consists of a domain marker followed by three discrete codes in quantization order. The domain marker distinguishes short video, e-commerce, advertising, and live streaming, while the three codes correspond to the outputs of the successive residual quantization levels.

OneReason adds special tokens for the domain markers and quantization codes to the language model's vocabulary. SIDs represent items in interaction histories and recommendation targets, and are combined with natural language to form training sequences. Subsequent semantic alignment training teaches the model the content associated with each SID, enabling it to refer to or generate these codes in user-understanding and recommendation tasks. The challenge provides item SIDs and their mapping tables for participants to use in model training.

\subsection{Multi-Granularity Semantic Alignment Pretraining}
\label{sec:semantic-alignment}

After item's Semantic IDs are added to the vocabulary, the model must learn both their meanings and the relationships between items. OneReason constructs pretraining data at four levels, token, item, relation, and user, to incorporate content semantics into behavioral modeling.

\paragraph{Token level.}
Token-level data establishes correspondences between individual codes or their combinations and natural language. Tasks include explaining the semantics of individual codes, mapping between code prefixes and descriptions in both directions, and composing a complete item description from explanations of constituent codes. These tasks help the model learn the semantic structure within item codes.

\paragraph{Item level.}
Item-level data uses complete SIDs for description generation, reverse retrieval, and content question answering. The model generates descriptions from SIDs, generates SIDs from descriptions, and answers questions about content attributes and intended audiences. Because compact discrete codes have limited information capacity, OneReason controls description granularity: it retains distinguishing information such as topics, categories, and brands while reducing details that are difficult to recover from the codes. This reduces inconsistencies between the supervision and the information carried by item representations.

\paragraph{Relation level.}
Relation-level data combines collaborative behavioral signals with semantic explanations. OneReason constructs item relations from signals such as searches following video views, transitions between items, and co-occurrence across domains, and generates text explaining these relations. For example, a transition from watching a video to searching for a product may reflect a shift from learning about a topic to seeking related products. Training sequences interleave item codes and relational explanations so that the model learns both behavioral associations and semantic connections.

\paragraph{User level.}
User-level data places items and their relations within user histories. One type organizes question--answer pairs by domain, conditioning predictions in a target domain on behavior in other domains. Another interleaves records from multiple domains chronologically and replaces some item codes with content descriptions. The former trains cross-domain conditional modeling; the latter preserves temporal order, allowing the model to interpret user interests using both content and timing.

Recommendation corpora are mixed with general text, mathematics, code, and multimodal data during pretraining~\cite{team2025kwai, yang2025kwai}. Training proceeds in three stages. First, the original model parameters are frozen while the embeddings and corresponding output-layer parameters for the new item tokens are trained. Next, full-parameter training learns recommendation semantics and behavioral patterns. Finally, the context length of training examples is extended to support longer user histories. The three stages use 110B, 449B, and 19B tokens, respectively.

This pretraining provides a foundation for modeling item semantics and user behavior. To give participants greater scope for exploration, the challenge leaves the design of supervised fine-tuning and reinforcement learning to participants and provides user interactions, item semantics, and SFT data. Participants can organize their own training data, construct reasoning supervision, and design rewards on top of this base model to study how post-training methods affect item understanding, user understanding, recommendation, and general knowledge.

\begin{table}[t!]
\centering
\caption{SFT data statistics.}
\label{tab:sft-data}
\small
\setlength{\tabcolsep}{4pt}
\begin{tabular*}{\textwidth}{@{\extracolsep{\fill}}lrrrr@{}}
\toprule
Task & Examples & Input tokens & Response tokens & \makecell[r]{Mean length\\(tokens)} \\
\midrule
\makecell[l]{Item understanding:\\ \texttt{nothink}} & 355,850 & 31,195,370 & 20,697,978 & 145.8 \\
\makecell[l]{Item understanding:\\ \texttt{think}} & 355,830 & 31,066,716 & 33,319,416 & 180.9 \\
\makecell[l]{User understanding:\\ \texttt{nothink}} & 24,509 & 108,083,890 & 3,535,879 & 4,554.2 \\
\makecell[l]{User understanding:\\ \texttt{think}} & 8,339 & 38,402,396 & 8,403,137 & 5,612.8 \\
Recommendation: \texttt{think} & 48,269 & 56,091,301 & 32,703,591 & 1,839.6 \\
General-Knowledge tasks & 152,005 & 169,034,420 & 1,047,291,417 & 8,001.9 \\
\midrule
Total & 944,802 & 433,874,093 & 1,145,951,418 & 1,672.1 \\
\bottomrule
\end{tabular*}
\end{table}

\section{Data Resources}
\label{sec:data-resources}

The challenge provides three types of data: supervised fine-tuning seed examples, user interactions, and item information.
Specifically, we provide 944,802 SFT seed examples, 500,000 anonymized user multi-domain history logs, and Semantic IDs/Caption alignment data for 35,914,095 items.

\subsection{Supervised Fine-Tuning Data}
\label{sec:sft-data}

Supervised fine-tuning (SFT) data~\footnote{OneReason SFT data: \url{https://huggingface.co/datasets/OpenOneRec/OneReasonSFT}} is provided as conversations consisting of task instructions and target responses, covering item understanding, user understanding, recommendation, and general-purpose tasks. Item understanding includes bidirectional mappings between descriptions and SIDs. User understanding includes relevant-behavior selection and interest-evolution analysis. Recommendation tasks generate target-domain SIDs from user histories, while general-purpose data includes instruction examples such as knowledge question answering.

\begin{figure}[t!]
\centering
\includegraphics[width=\textwidth]{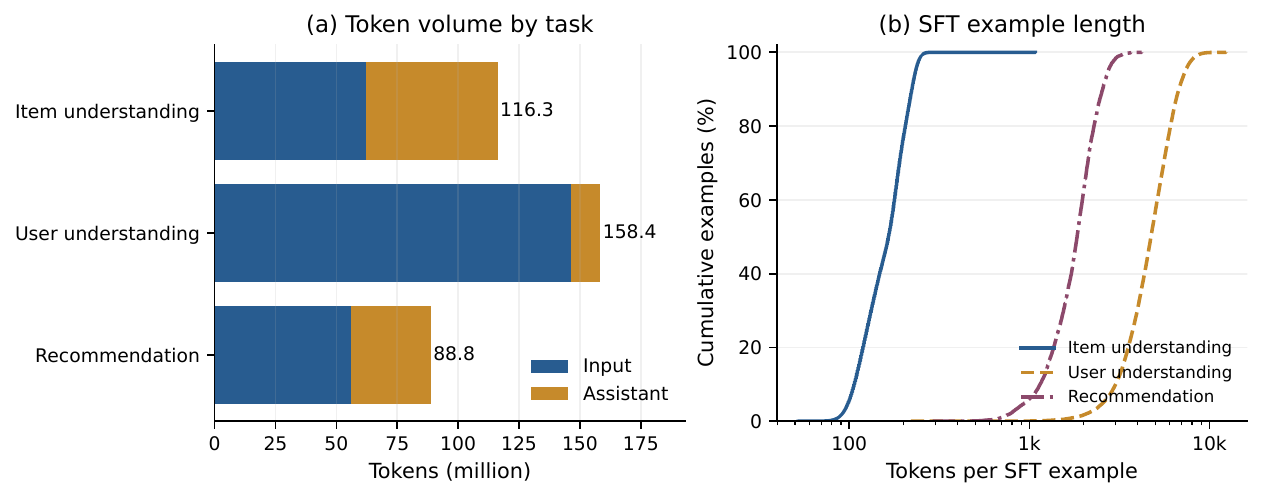}
\caption{Input and response token volumes (left) and cumulative distributions of example length (right) for recommendation-related SFT tasks.}
\label{fig:sft-data}
\end{figure}

Training examples use two response formats, \texttt{think} and \texttt{nothink}, which provide reasoning followed by an answer and a direct answer, respectively. Table~\ref{tab:sft-data} reports example counts, token counts, and average lengths by task. Figure~\ref{fig:sft-data} presents token volumes and example-length distributions for recommendation-related tasks.

In the preliminary round, we also provided item caption lists~\footnote{Partial SFT data Caption Sequence: \url{https://huggingface.co/datasets/OpenOneRec/Explorer_LLM_Rec_Competition/tree/main/SFT}} for a subset of the recommendation SFT examples. Each list contains the textual descriptions of items in the SID sequence, using the same descriptions that were used to construct the example's CoT. These lists were made available for participants reference .

\subsection{User Interactions}
\label{sec:user-interactions}

User interaction data covers short video, e-commerce, advertising, and live streaming. Historical sequences and interactions from the target period are organized by anonymized user~\footnote{Interaction Sequence: \url{https://huggingface.co/datasets/OpenOneRec/Explorer_LLM_Rec_Competition/tree/main/data}}. Each domain includes item hash identifiers, timestamps, and relevant feedback fields, with interactions across domains stored in the same user record.

The feedback fields reflect the interaction types in each domain. Short-video records include watch duration, completion, and engagement such as likes and comments. E-commerce records include impressions, clicks, cart additions, and purchases. Advertising records include clicks, conversions, and industry information. Live-streaming records include watch duration and counts of likes, comments, follows, and other interactions.

\begin{figure}[htbp]
\centering
\includegraphics[width=\textwidth]{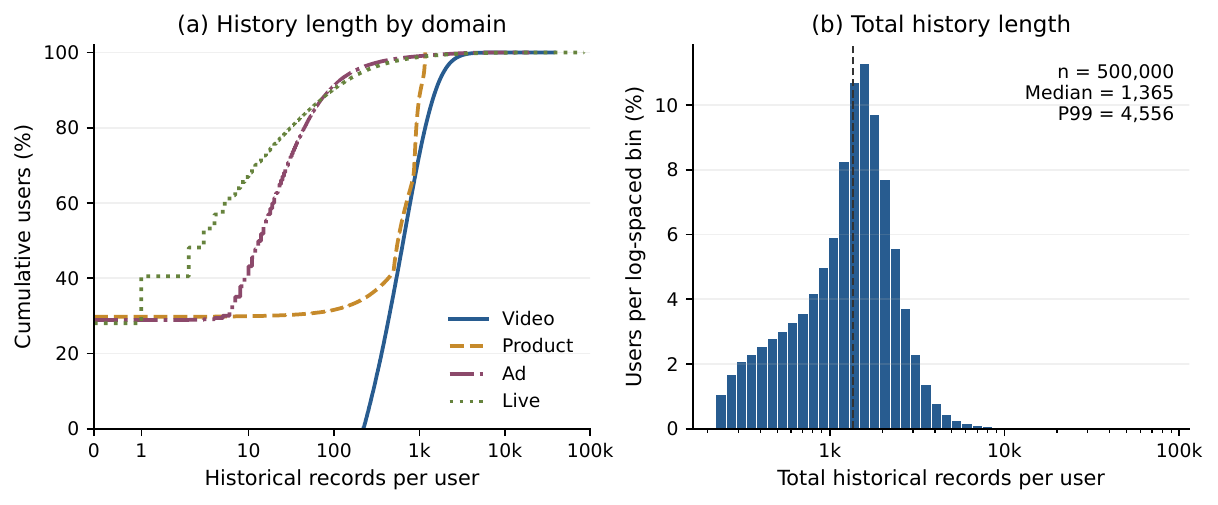}
\caption{Cumulative distributions of history length by domain (left) and the distribution of total history length across domains (right).}
\label{fig:user-history}
\end{figure}

Table~\ref{tab:user-history} summarizes user coverage, historical interaction counts, and sequence lengths by domain. Figure~\ref{fig:user-history} shows the corresponding length distributions: the left sub-figure presents cumulative distributions for individual domains, and the right sub-figure shows the distribution of total history length across domains for each user.

\begin{table}[t!]
\centering
\caption{Historical interaction statistics for 500,000 users.}
\label{tab:user-history}
\small
\setlength{\tabcolsep}{4pt}
\begin{tabular*}{\textwidth}{@{\extracolsep{\fill}}lrrrrr@{}}
\toprule
Domain & Users & Historical records & Mean length & Median length & P95 \\
\midrule
Short video & 500,000 & 409,416,641 & 818.8 & 649 & 1,950 \\
E-commerce & 351,337 & 258,269,396 & 516.5 & 562 & 1,137 \\
Advertising & 355,463 & 29,684,523 & 59.4 & 14 & 170 \\
Live streaming & 359,669 & 35,580,635 & 71.2 & 3 & 226 \\
\midrule
Total & 500,000 & 732,951,195 & 1,465.9 & 1,365 & 2,988 \\
\bottomrule
\end{tabular*}
\end{table}

\subsection{Item Semantics Alignment}
\label{sec:item-semantics}

Item semantics describes the topics, attributes, and categories of recommendation objects such as videos, products, advertisements, and live streams. The challenge provides this information in three forms: semantic IDs (SIDs), textual descriptions (captions), and three-level category labels. Each SID consists of a domain marker and three discrete codes; captions describe item content in natural language; and category labels specify a hierarchy from broad to specific categories. These resources are provided as mapping tables and linked to user interactions through the domain and anonymized item identifier (PID).

Table~\ref{tab:item-semantics} summarizes the numbers of items and semantic codes, along with the scale of textual descriptions in each domain. PIDs and SIDs do not have a one-to-one correspondence: multiple items may share the same SID. The shared SID percentage is the proportion of SIDs associated with multiple PIDs within a domain.

\begin{table}[htbp]
\centering
\caption{Item semantics statistics by domain.}
\label{tab:item-semantics}
\small
\setlength{\tabcolsep}{4pt}
\begin{tabular*}{\textwidth}{@{\extracolsep{\fill}}lrrrrr@{}}
\toprule
Domain & PIDs & SIDs & \makecell[r]{Caption\\coverage} & Caption tokens & \makecell[r]{Shared\\SIDs} \\
\midrule
Short video & 17,662,191 & 10,345,920 & 55.20\% & 1,653,348,287 & 20.84\% \\
E-commerce & 16,087,726 & 10,295,563 & 60.73\% & 198,488,381 & 20.20\% \\
Advertising & 2,056,889 & 1,069,596 & 69.10\% & 255,355,976 & 25.17\% \\
Live streaming & 107,289 & 62,070 & 99.69\% & 4,699,678 & 16.03\% \\
\bottomrule
\end{tabular*}
\end{table}

\section{Tasks and Evaluation}
\label{sec:tasks-evaluation}

\subsection{Overview of Evaluation Tasks}
\label{sec:evaluation-overview}

The challenge evaluates a single model across four capability dimensions: understanding item, understanding user, understanding recommendation, and understanding world~\footnote{Benchmark Data: \url{https://huggingface.co/datasets/OpenOneRec/OneReason-Eval-Benchmark} \\Benchmark Code: \url{https://github.com/Kuaishou-OneRec/OneReason_Eval_Benchmark}}. 
Item understanding requires reconstructing item content from SIDs. User understanding requires identifying historical interactions relevant to a specified topic and describing how the user's interests evolve. Recommendation requires generating items in a target domain from multi-domain histories. General knowledge is assessed through knowledge-based multiple-choice questions. Table~\ref{tab:evaluation-tasks} summarizes the inputs, outputs, and evaluation methods for the semi-final tasks.

\begin{table}[htbp]
\centering
\caption{Task inputs, model outputs, and evaluation methods across four capability dimensions.}
\label{tab:evaluation-tasks}
\footnotesize
\setlength{\tabcolsep}{3pt}
\renewcommand{\arraystretch}{1.15}
\begin{tabular}{@{}>{\raggedright\arraybackslash}p{.12\textwidth}>{\raggedright\arraybackslash}p{.14\textwidth}>{\raggedright\arraybackslash}p{.17\textwidth}>{\raggedright\arraybackslash}p{.20\textwidth}>{\raggedright\arraybackslash}p{\dimexpr.37\textwidth-24pt\relax}@{}}
\toprule
Capability & Task & Input & Model output & Evaluation \\
\midrule
Understanding Item & Item description generation (non-think) & An item's SID & A description of the item & Match information points in generated and reference descriptions; compute doubly weighted F1 using information importance and match quality. \\
\addlinespace
Understanding User& Relevant-behavior selection (non-think) & User history and an interest topic & A list of topic-relevant interactions selected from the history & Compare selected interactions with the reference answer; use F1 to measure selection precision and completeness. \\
\addlinespace
Understanding User & Interest-evolution chain generation (non-think) & User history and an interest topic & A chronological interaction chain with explanations of interest changes at each stage & Align events in order and combine interaction matching with explanation similarity to compute the Action--Logic Score. \\
\addlinespace
Understanding Recommendation & Cross-domain recommendation (think + non-think) & Multi-domain user history and a target domain & Candidate item SIDs in the target domain & Map and dedup candidate SIDs to PIDs based on compute Pass@32 in think and non-think mode separately.\\
\addlinespace
Understanding World & Knowledge question answering (non-think) & A common-sense knowledge question with A/B/C/D answers candidate & An choice letter & Compare with the reference answer and compute the proportion of correctly answered questions. \\
\bottomrule
\end{tabular}
\end{table}


\subsection{Test Data Construction}
\label{sec:test-data-construction}

Evaluation data is constructed from item content, user interaction logs, and knowledge questions. Item-understanding examples are organized by item, while user-understanding examples are organized by user history and interest topic. Recommendation examples use subsequent interactions in a specified domain as prediction targets. Each general-knowledge example is a multiple-choice question with a single correct answer. Table~\ref{tab:test-data} reports the evaluation dataset sizes.

\begin{table}[htbp]
\centering
\caption{Test examples by task.}
\label{tab:test-data}
\small
\setlength{\tabcolsep}{4pt}
\begin{tabular*}{\textwidth}{@{\extracolsep{\fill}}lrrrrr@{}}
\toprule
\multirow{2}{*}{Task} & \multicolumn{4}{c}{Sub-domain} & \multirow{2}{*}{Total} \\
\cmidrule(lr){2-5} & \makecell[c]{Short\\video} & E-commerce & Advertising  & \makecell[c]{Live\\streaming} & \\
\midrule
Item description generation & 574 & 412 & 624 & 496 & 2,106 \\
Relevant-behavior selection & --- & --- & --- & --- & 1,739 \\
Interest-evolution chain generation & --- & --- & --- & --- & 905 \\
Cross-domain recommendation & 10,285 & 1,000 & 1,000 & 1,000 & 13,285 \\
General knowledge question answering & --- & --- & --- & --- & 807 \\
\midrule
Total & --- & --- & --- & --- & 18,842 \\
\bottomrule
\end{tabular*}
\end{table}

\paragraph{Item understanding.}
Each example pairs an item SID with a reference description. OneReason uses multimodal large language models to generate descriptions from the original item content, followed by model-based review and manual quality checks. Descriptions emphasize different aspects across domains: video descriptions cover topics, people, settings, and major events; product descriptions cover product types, appearance, and attributes; advertisement descriptions cover promoted products or services, presented content, and marketing information; and live-streaming descriptions cover streamer appearance, content focus, audience, and interaction style.

Each domain also provides annotations that decompose reference descriptions into factual statements, assigning each statement an importance weight from 1 to 5. For example, product type and primary use are core information, while details such as color, decoration, and background are weighted according to their importance for identifying the content. During evaluation, the model generates a description from the SID, which is then semantically matched against these reference information points.

\paragraph{User understanding.}
Examples are built from chronological interaction histories spanning multiple domains. The histories retain dates, interaction types, and content. Videos, products, advertisements, and live streams are represented by SIDs, while searches retain their query text. The model can therefore use temporal order, interaction types, and content together to identify evidence relevant to a given interest topic.

To construct interest-evolution data, OneReason first organizes multi-domain logs into timelines and uses a large language model to extract candidate interest chains. These chains are then filtered using rules, model-based review, and manual checks. Filtering examines whether events follow chronological order, subsequent interactions introduce new needs or constraints, explanations are supported by behavioral evidence, and adjacent events exhibit clear progression or a shift in interest. Chains that merely group similar content, repeat similar interactions, or connect distant events without intervening links are removed.

The challenge defines two output formats for user understanding. Relevant-behavior selection takes a history and an interest topic and returns a JSON array of relevant interaction content. Interest-evolution chain generation organizes events around the topic, with each event containing a date (\texttt{date}), interactions (\texttt{action}), and an explanation (\texttt{logic}). The prompts distinguish three forms of evolution: completing needs within a scenario, causal progression of interests, and increasingly specific needs. Chains are limited to five steps; similar interactions on the same day and within the same stage of evolution may be merged into one event. Reference answers retain the behavioral evidence and explanation for each event, supporting subsequent action alignment and logic scoring.

\paragraph{Recommendation.}
Recommendation examples use user histories as inputs and subsequent interactions in the target domain as answers. Inputs are organized by domain and feedback type, including video viewing and engagement, product browsing and purchases, advertisement clicks, and live-stream viewing, follows, and tipping. Each example specifies one target domain and stores the SIDs and item identifiers of one or more ground-truth targets.

OneReason defines high-value targets separately for each domain: product clicks in e-commerce; video interactions whose watch duration exceeds the 75th percentile within the corresponding video-duration bucket; conversions such as activations and payments in advertising; and first-time tipping in live streaming. Construction further includes anomalous-example filtering, downsampling of popular items, and category balancing . The competition test set uses multi-domain histories as inputs, but each example specifies exactly one prediction domain: short video, e-commerce, advertising, or live streaming. An example may have multiple ground-truth interaction targets within that domain, and these items jointly form its reference answer for evaluation .

\paragraph{General knowledge.}
The test set consists of Chinese multiple-choice questions with a single correct answer. Each example contains a question, answer options, and a reference answer, presented through a standardized question--answer template. The prompt asks the model to return the option letter in the form ``The correct answer is \ldots'' in Chinese. This task does not depend on user histories or item SIDs and is evaluated separately from recommendation-related tasks .

\subsection{Evaluation Methods and Metrics}
\label{sec:evaluation-metrics}

The challenge uses a multi-task evaluation scheme. Each subtask is scored using its own metric, and the scores are combined according to task weights to obtain the overall score used for ranking participants. The evaluation procedures and metric definitions are described below.

\paragraph{Item understanding.}
Item descriptions are scored in three steps: information-point extraction, semantic matching, and weighted aggregation. A judge model decomposes each generated description into information points and matches them against the reference points. BERTScore F1 measures the quality of each matched pair. Uncovered reference information is counted as missing content, while unmatched generated information is counted as additional content. Both information importance and match quality contribute to the score.

For example $i$, let $g_k$ and $w_k$ denote a reference information point and its weight, and let $\hat g_j$ and $\hat w_j$ denote a generated information point and its weight. Let $M_i$ be the set of matched pairs, and $U_i$ and $\hat U_i$ the sets of unmatched reference and generated points, respectively. Match quality $q_{kj}\in[0,1]$ is computed using BERTScore F1 with \texttt{bert-base-chinese}~\cite{vaswani2017attention}. The weighted amounts of correctly matched, missing, and additional information are
\begin{equation}
\begin{aligned}
\mathrm{TP}_i &= \sum_{(k,j)\in M_i} w_k q_{kj},\\
\mathrm{FN}_i &= \sum_{(k,j)\in M_i} w_k(1-q_{kj})+\sum_{k\in U_i}w_k,\\
\mathrm{FP}_i &= \sum_{(k,j)\in M_i}\hat w_j(1-q_{kj})+\sum_{j\in\hat U_i}\hat w_j.
\end{aligned}
\end{equation}

The doubly weighted F1 for an example and the test-set score are
\begin{equation}
F_i^{\mathrm{item}}=
\frac{2\mathrm{TP}_i}{2\mathrm{TP}_i+\mathrm{FP}_i+\mathrm{FN}_i},
\qquad
S_{\mathrm{item}}=\frac{1}{N_{\mathrm{item}}}\sum_{i=1}^{N_{\mathrm{item}}}F_i^{\mathrm{item}}.
\end{equation}

This calculation accounts for the accuracy of core information, omissions from the reference content, and unsupported details introduced by the generated description.

\paragraph{User understanding.} We conduct two tasks to evaluate model's ability to extract user interest chain.
\subparagraph{Relevant-behavior selection.}
Relevant-behavior selection compares the sets of interactions in the model output and the reference answer. For example $i$, let $\hat A_i$ be the predicted set and $A_i$ the reference set. Precision, recall, and F1 are defined as
\begin{equation}
P_i=\frac{|\hat A_i\cap A_i|}{|\hat A_i|},
\qquad
R_i=\frac{|\hat A_i\cap A_i|}{|A_i|},
\qquad
F_i^{\mathrm{select}}=
\frac{2|\hat A_i\cap A_i|}{|\hat A_i|+|A_i|}.
\end{equation}

Precision measures the proportion of selected interactions that are relevant to the topic, while recall measures the proportion of reference interactions selected. The task score is the mean F1 across examples:
\begin{equation}
S_{\mathrm{select}}=\frac{1}{N_{\mathrm{select}}}
\sum_{i=1}^{N_{\mathrm{select}}}F_i^{\mathrm{select}}.
\end{equation}

\subparagraph{Interest-evolution chain generation.}
Interest-evolution chains are evaluated on both event selection and logical explanations. Let the reference chain contain $n$ events, $E=(e_1,\ldots,e_n)$, and the generated chain contain $m$ events, $\hat E=(\hat e_1,\ldots,\hat e_m)$. Each event consists of an interaction set and an explanation, denoted by $e_k=(A_k,l_k)$. Evaluation first aligns events across the two chains and then compares the explanations of aligned events .

Event alignment uses interaction-set F1 as the similarity measure:
\begin{equation}
a_{kj}=\frac{2|A_k\cap\hat A_j|}{|A_k|+|\hat A_j|}.
\end{equation}

Among one-to-one matchings that preserve event order in both chains, the matching with the largest total similarity is selected:
\begin{equation}
M^*=\underset{M\in\mathcal M_{\mathrm{mono}}}{\arg\max}
\sum_{(k,j)\in M}a_{kj}.
\end{equation}

Here, $\mathcal M_{\mathrm{mono}}$ denotes matchings in which both the reference-event indices and generated-event indices are strictly increasing. Events placed too early, omitted, or added therefore affect the chain score. Let $Q=\sum_{(k,j)\in M^*}a_{kj}$. The action-alignment scores are
\begin{equation}
P_{\mathrm{act}}=\frac{Q}{m},\qquad
R_{\mathrm{act}}=\frac{Q}{n},\qquad
F_{\mathrm{act}}=\frac{2Q}{m+n}.
\end{equation}

Logic alignment uses the same event matching. For each pair of matched explanations $l_k$ and $\hat l_j$, similarity is the equally weighted mean of Token F1 and ROUGE-L F1:
\begin{equation}
s_{kj}=\frac{1}{2}\operatorname{TokenF1}(l_k,\hat l_j)
+\frac{1}{2}\operatorname{ROUGE\!\text{-}\!L}_{F1}(l_k,\hat l_j).
\end{equation}

Let $x$ and $y$ be the token sequences of the two explanations, $c_x(t)$ and $c_y(t)$ the occurrence counts of token $t$, and $\operatorname{LCS}(x,y)$ the length of their longest common subsequence. Then
\begin{equation}
\operatorname{TokenF1}(x,y)=
\frac{2\sum_t\min\{c_x(t),c_y(t)\}}{|x|+|y|},
\qquad
\operatorname{ROUGE\!\text{-}\!L}_{F1}(x,y)=
\frac{2\operatorname{LCS}(x,y)}{|x|+|y|}.
\end{equation}

Let $T=\sum_{(k,j)\in M^*}s_{kj}$. The logic-alignment score and final chain score are
\begin{equation}
F_{\mathrm{logic}}=\frac{2T}{m+n},
\qquad
S_{\mathrm{chain}}=
\frac{1}{2}F_{\mathrm{act}}+\frac{1}{2}F_{\mathrm{logic}}.
\end{equation}

When multiple responses are generated for an example, their chain scores are averaged first, followed by averaging across test examples. Responses that cannot be parsed into the required structure receive a score of zero and remain in the denominator.

\paragraph{Recommendation.}
Recommendation is evaluated in both \texttt{think} and \texttt{nothink} modes. The \texttt{think} mode generates an analysis followed by 32 candidate SIDs; the \texttt{nothink} mode directly generates 32 candidate SIDs. The two outputs are merged and deduplicated to form the final candidate set.

For example $i$, let $C_i^{\mathrm{think}}$ and $C_i^{\mathrm{nothink}}$ denote the candidate sets from the two modes. Then
\begin{equation}
C_i=C_i^{\mathrm{think}}\cup C_i^{\mathrm{nothink}},
\qquad |C_i|\leq 64.
\end{equation}

Evaluation maps candidate SIDs to item PIDs and compares them with the target PIDs from the user's subsequent observed interactions. Let $\hat Y_i$ be the mapped candidate PID set for example $i$ and $Y_i$ its ground-truth target PID set. PIDPass is defined as
\begin{equation}
\operatorname{PIDPass}=
\frac{1}{N}\sum_{i=1}^{N}
\mathbf{1}\!\left[\hat Y_i\cap Y_i\ne\varnothing\right].
\end{equation}

An example receives a score of one if any candidate PID matches any ground-truth target, and zero otherwise. PIDPass is the average of these scores and is computed separately for the short-video, e-commerce, advertising, and live-streaming target domains.

\paragraph{General knowledge.}
For general knowledge, the option letter is extracted from the model's response and compared with the reference answer. Let $\hat y_i$ be the predicted answer to question $i$ and $y_i$ the reference answer. Accuracy is
\begin{equation}
\operatorname{Accuracy}=
\frac{1}{N_{\mathrm{general}}}
\sum_{i=1}^{N_{\mathrm{general}}}
\mathbf{1}[\hat y_i=y_i].
\end{equation}

\section{Competition Platform and Fairness}
\label{sec:platform-reproducibility}

\subsection{Platform and Computational Resources}
\label{sec:platform-resources}

The challenge runs on Kuaishou's Vanchin platform~\footnote{\url{https://www.streamlake.ai/product/wanqing}}, which provides model training, management, and evaluation services. In the preliminary round, each team receives a weekly training quota of 0.3B tokens for experiments through the platform's fine-tuning service. Participants may also train using their own computing resources. In the semi-final round, excepted the unlimited tokens of training quota, each qualifying team receives a development server equipped with 4 $\times$ 80GB premier GPUs, allowing participants to configure their training environments and implement their own optimization methods. 
We provide baseline models and training code, and the platform supplies inference resources for evaluation in both rounds.

All submitted models are also evaluated on Vanchin. The platform loads the submitted model parameters and applies a common inference and scoring procedure, reporting both sub-task scores and the overall score and updating the leaderboard daily. Participants have flexibility to explore data and training methods, while evaluation conditions remain consistent across submissions.

\subsection{Reproducibility and Fairness}
\label{sec:reproducibility-fairness}

After the semi-final round, we reproduce each submitted training pipeline from OneReason-8B-pretrain using the team's training data, code, and configuration, with a time limit of 72 hours for the complete pipeline.
Snapshots of training data and teacher models are preserved before reproduction, and the review covers both the construction of participant-created datasets and the training of teacher models.
Each reproduced model is evaluated five times, and its mean score is compared with the leaderboard score.
Generalization is also assessed on additional test data drawn from the same distribution.

We check models and training data for duplicates across teams. Models are compared using MD5 digests. For training data, complete conversations are normalized and assigned SHA-256 fingerprints to identify exact duplicates independently of auxiliary identifiers such as sample IDs. Examples that exactly match publicly released competition data are excluded before computing the number of shared examples, overlap proportions, and Jaccard similarity between teams.

To further safeguard fairness, we cross-check submitted training data against the test data for potential leakage at three levels: user histories, item identifiers, and semantic pairs. For recommendation and relevant-behavior selection, each user history is represented as a set of unique SIDs, and Jaccard similarity is computed between training and test histories. For each test history, its maximum similarity to any publicly released training history serves as a baseline; matches exceeding this baseline are selected for further review. At the item level, we check for SIDs that appear in the test set but are absent from the publicly released data. At the semantic level, we examine whether participant-constructed SID--description pairs cover test items not covered by the publicly released semantic pairs. Flagged records are manually reviewed in conjunction with their data sources and construction procedures.

\section{Participation and Leading Approaches}
\label{sec:participation-approaches}

\subsection{Participation and Competition Results}
\label{sec:participation-results}

The challenge attracted 2,031 registered participants from nine countries and regions, forming 1,206 teams. Of these, 812 teams completed valid evaluations in the preliminary round, with 33,059 valid submissions in total. Fifty teams advanced to the semi-final round, which received 4,861 valid submissions. Rankings in each round were determined by overall performance across item understanding, user understanding, recommendation, and general knowledge.

Leading approaches improved multi-task performance through supervised data refinement, reasoning training, and preference optimization on the provided pretrained base model. Some relied primarily on supervised fine-tuning, improving data quality and task proportions, while others incorporated reinforcement learning or preference optimization. Their main methods are summarized below.

\subsection{Leading Participant Approaches}
\label{sec:leading-approaches}

\paragraph{Champion Solution: self-sampled reasoning and target-masked training.}
The winning approach combines model self-sampling with Target-SID-Masked RFT to construct and learn from recommendation reasoning data. Reasoning text and recommendation answers are sampled from model checkpoints within the same training run. Trajectories that hit the target SID are retained, with reasoning deduplicated and the number of trajectories capped for each request. For these additional examples, loss is computed only on the reasoning text, with the final SID loss masked; original SFT examples retain full response supervision.

This design arose from performance degradation when successful trajectories were added back into training. When multiple reasoning trajectories shared the same target and both the reasoning and final SID were supervised, candidate diversity in \texttt{think} decreased and its overlap with \texttt{nothink} increased, while diversity in \texttt{nothink} remained largely unchanged. Masking the target SID loss on the additional trajectories improved both candidate complementarity between the two modes and recommendation performance. Ablations further showed that this treatment outperformed full-trajectory supervision and merely reducing the weight of repeated answers.

Trajectory source and quantity also affected training. In comparisons using target masking, self-sampled trajectories yielded better recommendation results than data generated by a larger teacher model. Retaining a limited number of distinct trajectories was beneficial, but adding more reduced the gains. Candidate analysis also showed that reducing overlap between the two modes alone was insufficient to improve hit rates. These results support jointly controlling the source, quantity, and answer-supervision strength of additional reasoning trajectories in the challenge's combined \texttt{think} and \texttt{nothink} recommendation setting.

\paragraph{Runner-Up Solution: supervised data construction for multi-task learning.}
The runner-up approach improves multi-task SFT data through item-description rewriting and domain-specific recommendation supervision. Constrained rewriting preserves attributes supported by the original descriptions, including category, brand, material, specifications, and function, while removing marketing language and service information and prohibiting the addition of unsupported attributes. With other task data held fixed, this treatment improved item understanding in the short-video and e-commerce domains.

Inspection of recommendation data revealed that some reasoning texts included activities, such as searches, that were absent from the input history. The team further compared supervision formats across domains, ultimately retaining \texttt{think} for e-commerce and using \texttt{nothink} for short video, advertising, and live streaming. With other task data held fixed, this mixed configuration outperformed using either \texttt{think} or \texttt{nothink} across all four domains, indicating that a uniform reasoning-supervision format was unnecessary under these experimental conditions.

The final approach also incorporates knowledge-data refinements and retains the original user-understanding data. Changes to user-intent cues and interest-evolution expressions did not improve both user-understanding subtasks simultaneously. Restating knowledge answers also had different effects on general knowledge and some recommendation tasks.

\begin{figure}[t!]
\centering
\includegraphics[width=\textwidth]{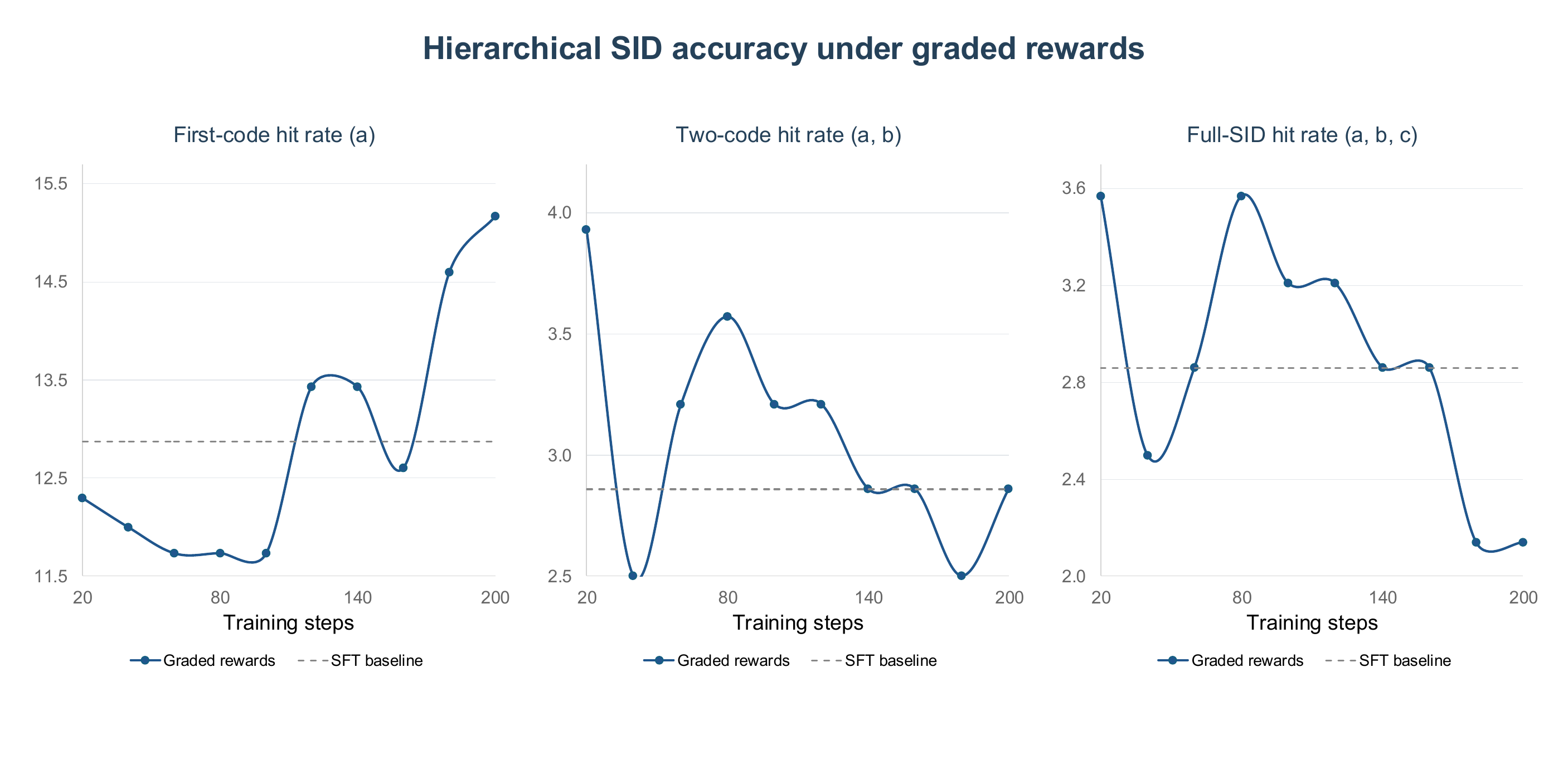}
\caption{Accuracy at successive SID levels under graded rewards.}
\label{fig:hierarchical-hits}
\end{figure}

\paragraph{Third-place Solution: staged post-training with task-specific feedback.}
The third-place approach proceeds through SFT, reinforcement learning for general knowledge, reinforcement learning for user understanding, and preference optimization for recommendation. During SFT, recommendation examples are constructed in both \texttt{think} and \texttt{nothink} formats, with input markers specifying the generation mode. To account for substantial differences in response length across tasks, training uses pack-level cross-entropy: losses are first averaged over valid response tokens within each packed sequence, then aggregated with equal weight across packed sequences. This prevents packs with more valid response tokens from receiving greater weight solely because of their token count, while preserving the relative weighting of long and short responses within each pack.

In general-knowledge reinforcement learning, rewarding only answer correctness led to repeated derivations and repeated answers. Requiring a single answer, adding a progressive length penalty, and selecting response groups with differing rewards reduced abnormal repetition and improved general-knowledge performance at this stage. The user-understanding stage uses behavior-selection F1 as feedback, allowing partially correct responses to receive rewards.

The recommendation stage uses direct preference optimization (DPO). Even with increased online sampling, many requests failed to generate any known interaction target, leaving all-zero reward groups without a relative learning signal. The team therefore constructs positive examples from observed interaction targets in the logs and selects negative items from the same domain, excluding other known positives and historical interactions. Negatives are organized using the SID hierarchy, including items with different first-level codes and items that share the first-level code but differ at the second level.

Each preference pair shares the same reasoning text and differs only in the final item SID, enabling comparison between recommendation outcomes. Training combines a label-smoothed preference loss with a positive-example generation loss. This stage improved recommendation performance in short video, e-commerce, and live streaming, while advertising performance remained largely stable.

\begin{figure}[t!]
\centering
\includegraphics[width=\textwidth]{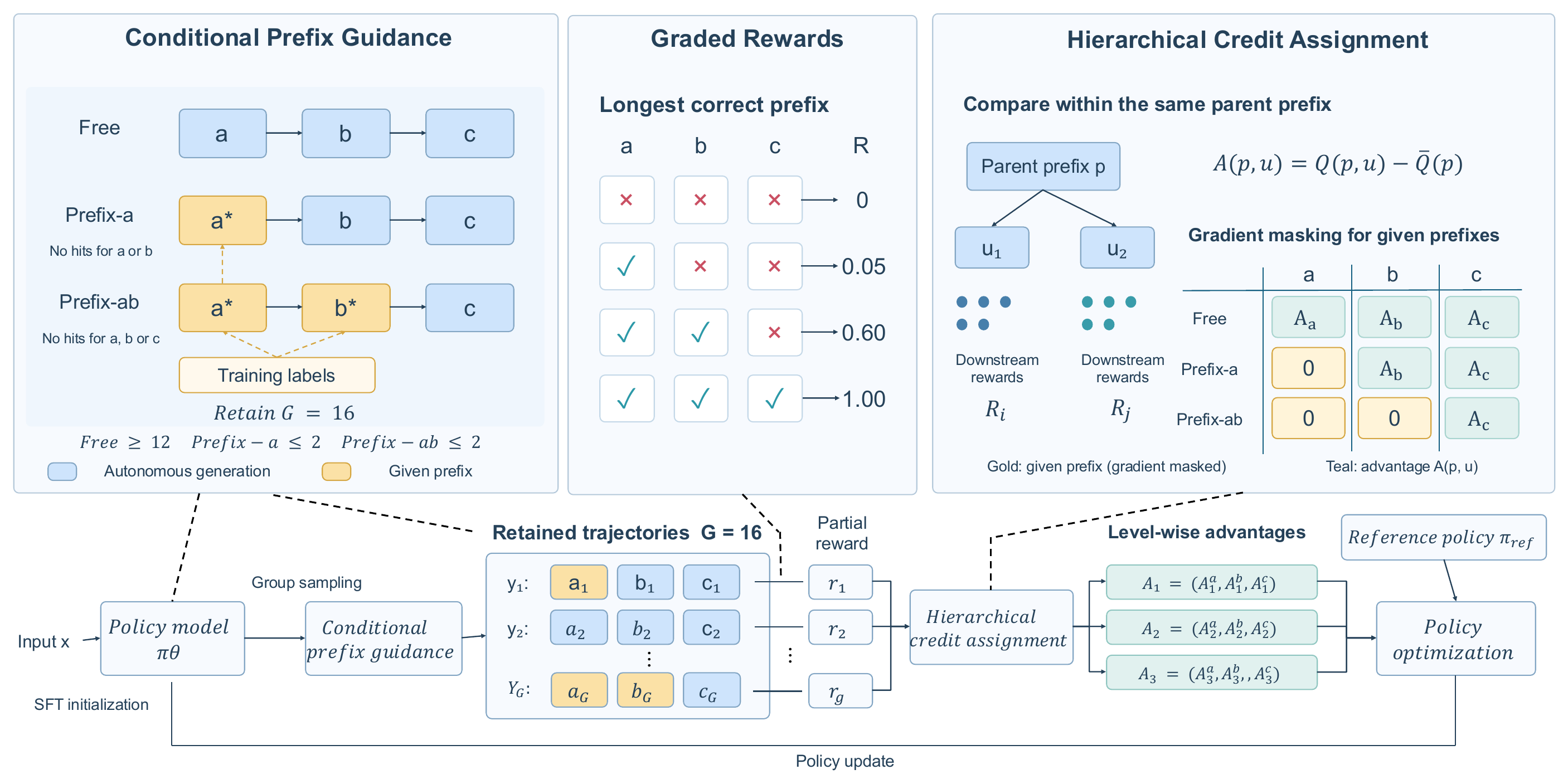}
\caption{Prefix guidance and hierarchical credit assignment.}
\label{fig:prefix-credit}
\end{figure}

\subparagraph{Innovative Solution: Prefix guidance and hierarchical credit assignment.}
The fourth-place team proposes prefix guidance and hierarchical credit assignment to provide training feedback for hierarchical SID generation. Applying group relative policy optimization (GRPO) with rewards only for complete SID hits leaves most sampling groups with identical rewards, making effective relative updates difficult. With graded rewards based on the number of consecutive correct codes, first-level accuracy rises during training without corresponding improvements in two-level or complete-SID accuracy (Figure~\ref{fig:hierarchical-hits}). The team interprets this pattern as a preference for popular first-level codes that readily earn shallow rewards. Another experiment injects correct answers directly into sampling groups, but increases in average reward mainly come from the injected answers, without corresponding improvements in autonomous generation.

Figure~\ref{fig:prefix-credit} illustrates the prefix-guided sampling procedure. The model first generates freely. If no candidate has both of the first two codes correct, the target's first code is supplied to guide generation of the remaining codes. If there is still no complete hit, the first two codes are supplied and the model generates the final code. Most trajectories in each group remain freely sampled, with the number of guided trajectories limited, increasing opportunities to learn later codes under correct prefixes.

Hierarchical credit assignment further separates the contribution of each code level to the reward. Within a request, the method compares only code choices with the same parent prefix, estimates the value of each choice from its mean downstream reward, and computes advantages relative to the group baseline for separate updates at each level. For the last two levels, only candidates with correct prefixes and autonomously generated current codes are included. Supplied prefixes and preceding segments are excluded from policy-gradient computation, so rewards attributable to supplied prefixes do not directly reinforce the model's generation of those prefixes.

\begin{figure}[htbp]
\centering
\includegraphics[width=\textwidth]{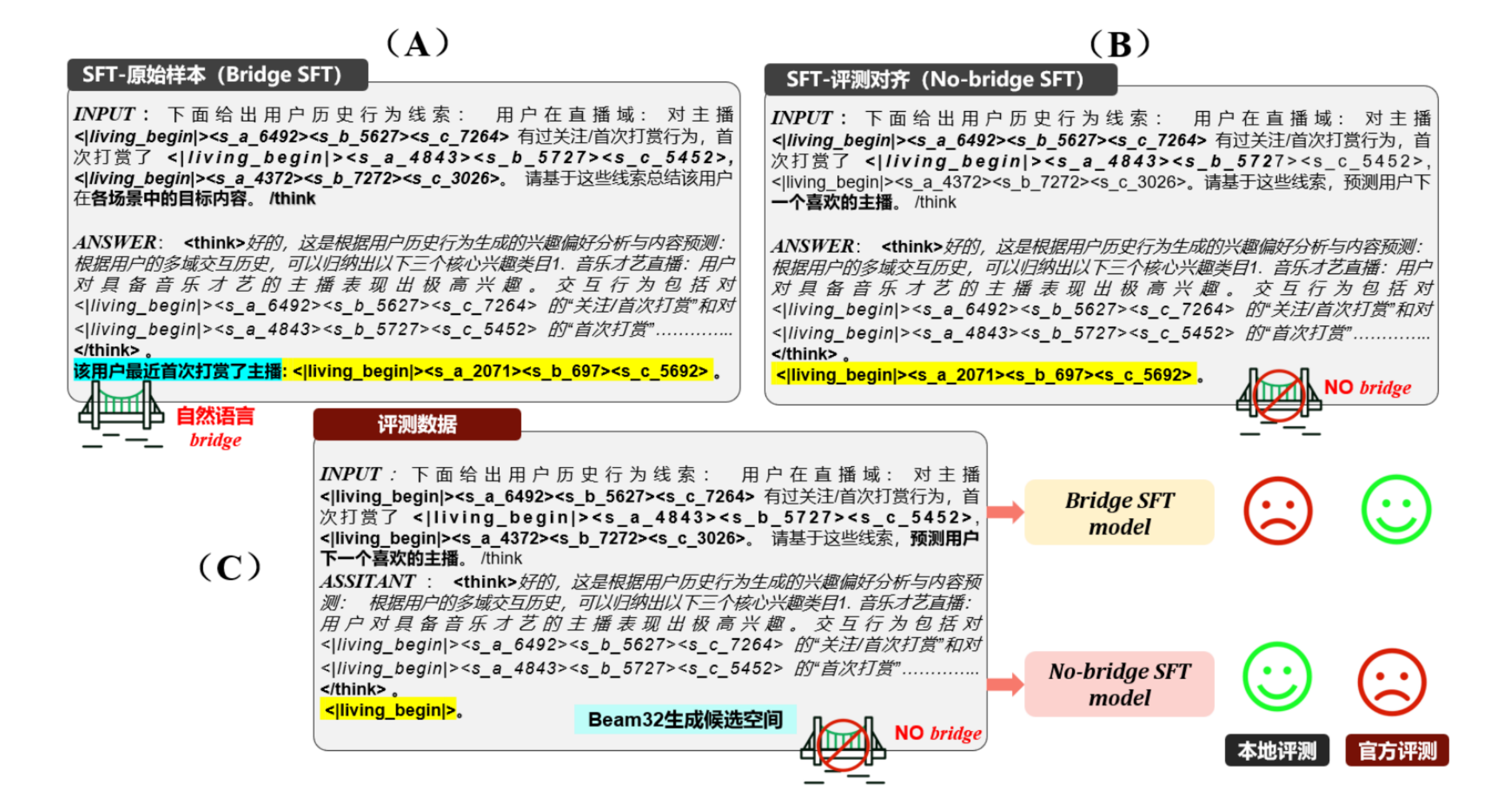}
\caption{The Bridge discrepancy between training and inference.}
\label{fig:bridge-phenomenon}
\end{figure}

\subparagraph{Innovative Solution: Recommendation generation with a natural-language transition.}
The fifteenth-place team examines the natural-language phrase between the reasoning text and the SID, termed a Bridge. Original SFT examples end the reasoning with a phrase such as ``The user recently tipped this streamer for the first time'' before producing the SID, whereas inference generates candidates directly from the domain marker. With the model, input, and reasoning held fixed, changing only whether the Bridge is retained substantially alters the candidate set and the top recommendation. Removing the Bridge from training examples to align the training and inference formats improved local validation performance but reduced online test performance (Figure~\ref{fig:bridge-phenomenon}).

Candidate analysis shows that models trained with a Bridge are more likely to reuse historical items. Removing it produces more candidates outside the history, without consistently increasing true-target hits. Moreover, reasoning-text training loss continues to decrease without a corresponding improvement in SID prediction. The team hypothesizes that the transition phrase may affect competition between historical items and targets outside the history within the candidate set, and that aligning output formats alone is insufficient to address this issue.

\begin{figure}[htbp]
\centering
\includegraphics[width=\textwidth]{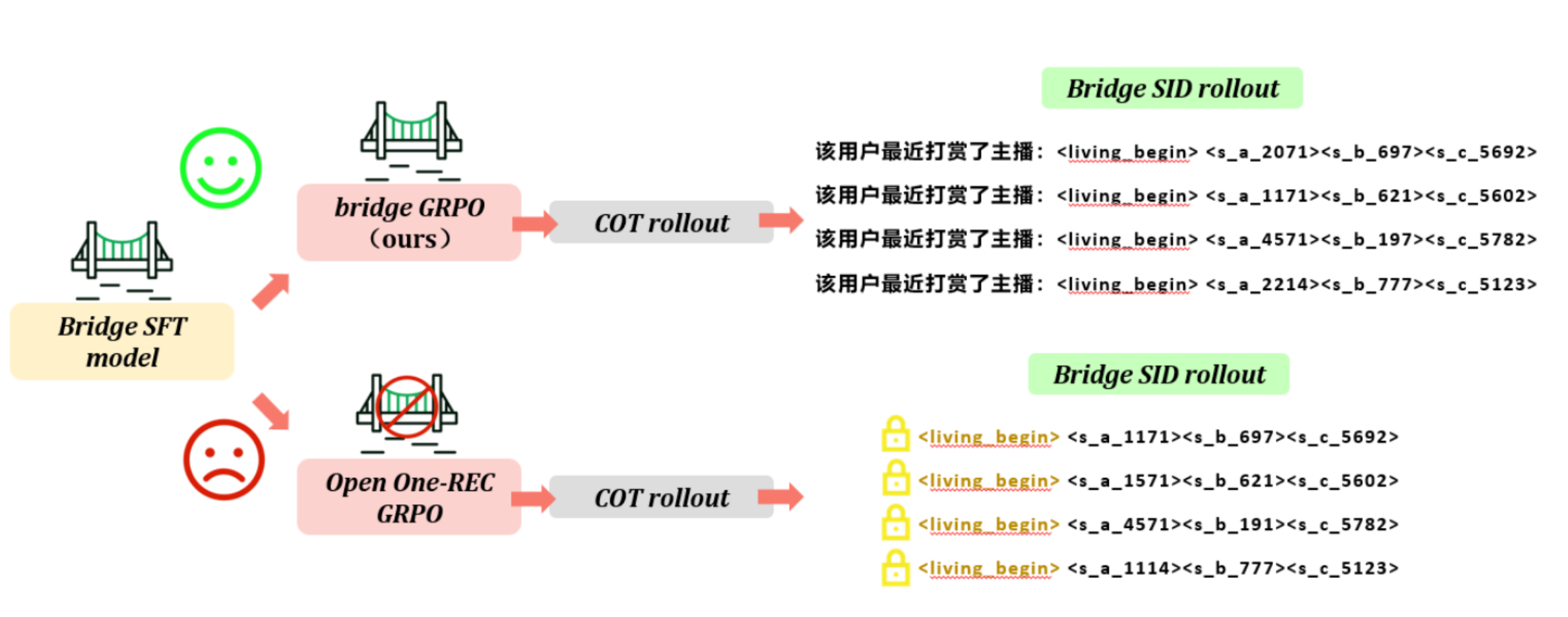}
\caption{The Bridge-GRPO training approach.}
\label{fig:bridge-framework}
\end{figure}

To learn this transition from recommendation feedback, the team proposes Bridge-GRPO, retaining a two-stage process of reasoning followed by answer generation. As shown in Figure~\ref{fig:bridge-framework}, OpenOneRec-GRPO fixes the domain prefix before generating the SID in the answer stage. Bridge-GRPO removes this constraint, allowing the model to generate a natural-language transition and the subsequent SID autonomously, and optimizes this generation process using recommendation rewards.

\section{Conclusion}
In this report, we have provided a comprehensive review of the KUAISHOU Explorer LLM-Rec Challenge 2026, including its motivation, task design, OneReason foundation, opensource data resources, evaluation benchmark, reproducibility and fairness.
We have also reviewed the leading participant solutions and summarized the key insights that emerged from the competition.
Throughout the challenge, participants developed many valuable approaches to semantic alignment, reasoning supervision, multi-task learning, reinforcement learning, preference optimization, and hierarchical recommendation, demonstrating both the promise of reasoning generative recommendation and the challenges that remain.

We believe that making realistic data and rigorous evaluation resources publicly available is essential for collective progress in recommendation foundation model research.
Therefore, in this report, we fully open-source the challenge data and benchmark, including the multi-domain user interactions, item semantics, task-formatted SFT data, and evaluation benchmarks.
We hope these resources will enable reproducible comparisons, support the exploration of new methods, and provide a shared foundation for the research community.

Finally, we believe that connecting large language models with recommender systems is a highly promising direction.
This challenge represents only a small step, and we hope that more researchers will join this wave, contribute new ideas and technologies, and work together to shape the future of recommender systems.

\printbibliography

\newpage
\quad \\
\quad \\

\newpage
\section{Author List and Acknowledgement}

\noindent
\textbf{Core Organizers}\quad
Jiangxia Cao\textsuperscript{$\dagger$},
Hao Peng,
Wenlong Xu,
Jiaxin Deng,
Zhixin Ling,
Xingmei Wang,
Kun Shang,
Can Tang,
Zhihuai Cai,
Jun Du,
Fang Su,
Xiaojuan Liu,
Yiling Li,
Chenglong Yu,
Chongling Rao,
Haixuan Gao,
Haitao Xu,
Jian Liang,
Ruiming Tang

\vspace{0.5em}
\noindent
\textbf{Organizers}\quad
Chenglong Chu,
Guohong Mu,
Honghui Bao,
Hui Wang,
Jialong Chen,
Jiao Ou,
Muhao Wei,
Peng Zhang,
Renpu Liu,
Ruochen Yang,
Shugui Liu,
Xinqi Jin,
Yan Sun,
Yifan Wang,
Yingzhi He,
Yufei Ye,
Yusen Huo

\vspace{0.5em}
\noindent
\textbf{Competition Finalists}\quad
Tingkuo Wang,
Jihong Zhang,
Lanxi Zhu,
Pengyuan Liu,
Zhipeng Yi,
Luankang Zhang,
Hang Lv,
Xuyang Zhi,
Tianyu Li,
Bintao Wu,
Chuang Ou,
Siyue Su,
Ziyuan Wang,
Yuliang Sun,
Baiyan Che,
Feiyang Xu,
Shiwen Zhang,
Shiteng Cao,
Chongcong Jiang,
Yuan Fang,
Xiangwu Yang,
Hao Deng,
Zijian Du,
Pengxun Wang,
Xiaoming Wang,
Shun Qin,
Yingqi Song,
Tianyi Li,
Naixiao Peng,
Chenyu Zhou,
Qiliang Jiang,
Quan Zheng,
Cheng Jin,
Siying Zeng,
Hongjia Xu,
Junwu Hu,
Teng Fu,
Zhengkang Mei,
Haijun Yu,
Kai Li,
Shengyang Zhou,
Zhijia Wei,
Siyi Xiong,
Bo Liu,
Zichun Guo,
Zhubin Han,
Jinpeng Fu,
Bingqian Liu,
Yuyi Wang,
Yu Liu,
Qinghai Tan,
Ruijie Zhou,
Zhuohang Li,
Zhijia Zhong,

\vspace{0.5em}
\noindent
\textbf{Competition Steering Committee}\quad
Xiangnan He,
Jirong Wen,
Min Zhang,
Wenwu Ou,
Peng Jiang,
Han Li,
Kaiqiao Zhan,
Yanan Niu,
Lantao Hu,
Kun Gai

\vspace{0.5em}
\noindent
{\textsuperscript{$\dagger$}Correspondence: caojiangxia@kuaishou.com}

\end{document}

%% file: commands.tex
\usepackage{xspace}

\newcommand{\cdashlinerow}[2]{%
  \cdashline{#1}%
  \noalign{\global\let\CT@row@color\relax\vskip0pt}%
  \rowcolor{#2}%
}

\newcommand{\eat}[1]{}

\crefformat{section}{\S#2#1#3}

